\documentclass{webofc}
\usepackage{scrextend}
\usepackage[varg]{txfonts}   
\usepackage{cprotect}
\usepackage{xcolor}
\usepackage[thinc]{esdiff}
\usepackage{amsmath}
\usepackage{tablefootnote}
\usepackage{booktabs}
\usepackage{hhline}
\usepackage{color}
\usepackage{dcolumn}
\usepackage{bm}
\usepackage{mathtools}
\usepackage{fancyvrb}
\usepackage{hyperref}
\usepackage{graphicx}
\usepackage{caption}
\usepackage{subcaption}
\usepackage{ragged2e}
\usepackage[normalem]{ulem}
\usepackage[font=scriptsize]{caption}

\newcommand\yst{\bgroup\markoverwith{\textcolor{purple}{\rule[0.5ex]{2pt}{0.4pt}}}\ULon}

\usepackage{soul}

\begin{document}
\title{\texttt{class\_sz} I: Overview}
%
%

\author{\lastname{B. Bolliet}\inst{1,2} \and
        \lastname{A. Kusiak}\inst{9} \and
        \lastname{F. McCarthy}\inst{1,2,3} \and
        \lastname{A. Sabyr}\inst{9} \and
         \lastname{K. Surrao}\inst{9}\and
         \lastname{J. C. Hill}\inst{3,9}\and
          \lastname{J. Chluba}\inst{8}\and
           \lastname{S. Ferraro}\inst{5,6}\and
            \lastname{B. Hadzhiyska}\inst{6}\and
              \lastname{D. Han}\inst{1,2}\and
              \lastname{J. F. Mac\'{\i}as-P\'erez}\inst{7}\and
               \lastname{M. Madhavacheril}\inst{13}\and
                \lastname{A. Maniyar}\inst{10}  \and
                \lastname{Y. Mehta}\inst{4}  \and
                \lastname{S. Pandey}\inst{9}  \and     
                \lastname{E. Schaan}\inst{10}  \and     
                \lastname{B. Sherwin}\inst{1,2} \and   
                 \lastname{A. Spurio Mancini}\inst{12}  \and    
                 \lastname{\'{I}. Zubeldia}\inst{1,11}           
}

\institute{\footnotesize Kavli Institute for Cosmology, University of Cambridge, Madingley Road, Cambridge CB3 0HA
\and
           DAMTP, Centre for Mathematical Sciences, Wilberforce Road, Cambridge CB3 0WA, UK
\and
           Center for Computational Astrophysics, Flatiron Institute, New York, NY, USA 10010
\and
            School of Earth and Space Exploration, Arizona State University, Tempe, AZ, USA 85287
\and  Lawrence Berkeley National Laboratory, One Cyclotron Road, Berkeley, CA, USA 94720
\and Berkeley Center for Cosmological Physics, Department of Physics, University of California, Berkeley, CA, USA 94720
\and Laboratoire de Physique Subatomique et de Cosmologie, Universit\'e
Grenoble-Alpes, CNRS/IN2P3, 53, 
avenue des Martyrs, 38026 Grenoble
cedex, France
\and Jodrell Bank Centre for Astrophysics, Alan Turing Building, University of Manchester, Manchester M13 9PL
\and Department of Physics, Columbia University, New York, NY, USA 10027
\and SLAC National Accelerator Laboratory 2575 Sand Hill Road Menlo Park, California 94025, USA
\and Institute of Astronomy, University of Cambridge, Madingley Road, Cambridge CB3 0HA
\and Mullard Space Science Laboratory, University College London, Dorking, RH5 6NT, UK
\and Department of Physics and Astronomy, University of
Pennsylvania, 209 South 33rd Street, Philadelphia, PA, USA 19104
          }

\abstract{%
   \texttt{class\_sz} is a versatile and robust code in C and Python that can compute theoretical predictions for a wide range of observables relevant to cross-survey science in the Stage IV era. The code is public at \href{https://github.com/CLASS-SZ/class_sz}{https://github.com/CLASS-SZ/class\_sz} along with a series of tutorial notebooks (\href{https://github.com/CLASS-SZ/notebooks}{https://github.com/CLASS-SZ/notebooks}). It will be presented in full detail in paper II. Here we give a brief overview of key features and usage. 
}
\maketitle
\section{Introduction}
\label{intro}
With \textit{Planck} and \textit{WMAP} full-sky observations of the Cosmic Microwave Background (CMB), one of the key numerical and theoretical challenges was to accurately predict the CMB angular anisotropy power spectra in temperature and polarisation, sourced by primordial universe physics, Big Bang Nucleosynthesis, recombination, reionisation, and the Universe's expansion over the full cosmic history. The first numerical code that incorporated all of the relevant physics while achieving an efficient computing time was CMBFAST in 1996 \cite{Zaldarriaga:1996xe}. It was followed by the release of two subsequent codes built on the same set of ideas as CMBFAST: \texttt{camb} in 2000 \cite{CAMB} and \texttt{class} in 2011 \cite{lesgourgues2011cosmic}. Since then, \texttt{camb} and \texttt{class} have been used in thousands of publications, including by the \textit{Planck} and \textit{WMAP} collaborations, and continue to be used in the analysis pipelines of ground based CMB observatories today (e.g., the Atacama Cosmology Telescope (ACT) and the South Pole Telescope (SPT)) and tomorrow (e.g., the Simons Observatory (SO) and CMB-S4). 

With ACT and SPT, as well as forthcoming SO and CMB-S4 low-noise and high-resolution multi-frequency maps of the extragalactic sky, a new scientific window is opening-up: we can now measure the CMB anisotropies at arcminute scales. Their properties are dictated by the interaction of CMB photons with matter, during the formation of the Large Scale Structure (LSS) of the Universe. The dominant effects responsible for this so-called secondary CMB anisotropies are weak lensing by dense structure forming regions and inverse-Compton scattering between CMB photons and hot electrons in the Intracluster Medium (ICM) and Circum Galactic Medium (CGM), i.e., the Sunyaev--Zeldovich (SZ) effect. Furthermore, at small scales, the  sky temperature anisotropy receives contributions from other sources, such as the Cosmic Infrared Background (CIB) light from star-forming galaxies and radio emission from active galactic nuclei, so that the CMB is generally a sub-dominant contribution. Thus, the small-scale temperature anisotropy is a complicated mixture sourced by different physical effects. 

Fortunately, we have a good  theoretical understanding of these physical effects. Within a cosmological model and with a relatively small number of astrophysical assumptions, we can precisely predict the small-scale temperature anisotropy. Moreover, using galaxy survey data, we can construct estimators based on statistical cross-correlations that can single out different physical processes. For instance, we can measure the cross-correlations between the SZ effect and galaxies at various redshifts, making a tomography of the ICM, which we can then use to test galaxy formation models, see, e.g., Ref. \cite{Battaglia:2019dew}. We can also use the cross-correlations between galaxy and CMB weak lensing to probe the growth of structure at low-$z$ and address the S8 tension (see, e.g., Ref. \cite{ACT:2023oei}). These scientific opportunities are huge and constitute one the main driving forces of the field in the post-\emph{Planck} era. 

Extracting scientific information from small-scale temperature anisotropy and cross-correlation measurements comes with new theoretical and computational challenges. There are two powerful theoretical frameworks that enable us to make models and predictions for the interpretation of this new data: perturbation theory (including standard perturbation theory and Effective Field Theory of LSS, e.g., Ref. \cite{Carrasco:2012cv}) and the halo model formalism (see, e.g., Refs. \cite{Cooray:2002dia,Philcox:2020rpe,Asgari:2023mej}). Computationally, their implementation is challenging because one has to solve high-dimensional integrals (e.g., over comoving volume, halo masses, wavenumbers etc.) relying on many ingredients (halo mass function, halo biases, loop integrals, radial profile of tracers etc.). \texttt{class\_sz} is one of the few numerical codes that solve these challenges simultaneously for a  maximum number of observables. Another notable attempt is \texttt{ccl}, developed within the Vera Rubin Observatory collaboration\footnote{\href{https://github.com/LSSTDESC/CCL}{https://github.com/LSSTDESC/CCL}} \cite{LSSTDarkEnergyScience:2018yem}. 

\section{Structure of \texttt{class\_sz}}
\label{sec-1}

\texttt{class\_sz} is built based on the Boltzmann code \texttt{class}; thus, everything that can be computed by \texttt{class} can be computed by \texttt{class\_sz}. In \texttt{class\_sz}, we have added three "modules" to the original \texttt{class} code. These are: 
\begin{itemize}
\item \texttt{class\_sz.c}: the core of the code, where all the important quantities are computed (e.g., power spectra, bispectra, halo mass function etc),
\item  \texttt{class\_sz\_tools.c}: less central functions, such as numerical routines for integration or interpolations,
\item  \texttt{class\_sz\_clustercounts.c}: SZ cluster number counts calculations.
\end{itemize}
The python wrapper of \texttt{class} is extended \ to \texttt{class\_sz}. We have added all required \texttt{class\_sz} functions in  \texttt{classy.pyx} to "cythonize" them. In addition, we created a submodule in the  \texttt{class\_szfast.py} file dedicated to interface neural network emulators with the C code (see \ref{sec-4} and \ref{sec-5}).

\section{General features}
\label{sec-2}

We have invested significant efforts to make the code as fast as possible. Notably, all large ``for loops'' are parallelized with OpenMP; integrals are either solved with an adaptive solver (borrowed from \texttt{CosmoTherm} \cite{Chluba:2011hw}) or with Fast Fourier Transforms with the fftw3 library and the FFTLog algorithm when relevant. For a number of cosmological models, including $\Lambda$ Cold Dark Matter (CDM), $w$CDM, and $\Sigma m_\nu+\Lambda$CDM, \texttt{class\_sz} can call high accuracy neural network emulators to predict the CMB $C_\ell$'s and matter power spectrum. As a rule of thumb, we aim for all our calculations (for a given observable) to be completed in less than 0.2s\footnote{Of course, this depends on scales and number of cores used.}. 

\section{Availability and tutorials}
\label{sec-3}
The code is available on GitHub \href{https://github.com/CLASS-SZ/class_sz}{https://github.com/CLASS-SZ/class\_sz}  where it is regularly updated. Installation instructions are given in the README file and documentation will soon be provided. Once installed, one can carry out calculations using the C executable or within python code using the python wrapper \texttt{classy\_sz}.  A Google colab notebook with example calculations is available online at \href{https://colab.research.google.com/drive/1AULgG4ZLLG1YXRI86L54-hpjWyl1X-8c?usp=sharing}{class\_sz\_colab.ipynb} and can be run from any internet browser. 
In addition, we have provided an extensive set of notebooks that show how to obtain most \texttt{classy\_sz} outputs, in a self-explanatory way. These notebooks are stored at: \href{https://github.com/CLASS-SZ/notebooks}{https://github.com/CLASS-SZ/notebooks}.

\section{Fast calculations of CMB power spectra and fast MCMC's}
\label{sec-4}
We have imported the high accuracy \texttt{cosmopower}\footnote{\href{https://github.com/alessiospuriomancini/cosmopower}{https://github.com/alessiospuriomancini/cosmopower}} emulators for CMB TT, TE, EE angular anisotropy power spectra developed in Ref. \cite{Bolliet:2023sst}, building on previous work from Ref. \cite{SpurioMancini:2021ppk}. This allows \texttt{class\_sz} to predict these spectra in less than 50 ms, compared to around a minute if the calculations were to be done with \texttt{class} for the same accuracy requirements. Our accuracy requirements are such that our results are converged to better than 0.03$\sigma$'s of CMB-S4 across the entire multipole range between $\ell=2$ and $10^{4}$. 

This means that we can run Markov Chains Monte Carlo (MCMC) analyses of CMB power spectra data and reach convergence within a few minutes, compared to $\approx1$ week if we were to run the analysis with \texttt{class} tuned at the same accuracy. See \href{https://github.com/CLASS-SZ/notebooks/blob/main/class_sz_tutorial.ipynb}{class\_sz\_tutorial.ipynb} and \href{https://github.com/CLASS-SZ/notebooks/tree/main/mcmcs/cobaya/notebooks}{cobaya notebooks}.

\section{Fast calculations of matter power spectrum}
\label{sec-5}
The  matter power spectrum is the fundamental building block for the modelling of all LSS observables. It is often computed by \texttt{camb} or \texttt{class}, which solve the density perturbation evolution equations.  We have also imported matter power spectra emulators (developed in Ref. \cite{Bolliet:2023sst} too), so we can completely bypass the perturbation module of \texttt{class} and obtain interpolators for the linear and non-linear matter power spectra covering redshifts between 0 and 5 and wavenumbers between $10^{-4}$ and $50$ Mpc$^{-1}$ in $\approx 0.1$s compared to minutes for the equivalent  \texttt{class} calculation. Note that currently our non-linear prescription is the fiducial \texttt{hmcode} \cite{Mead_2021} calculation (i.e., dark-matter only) but we are planning to interface a wide range of power spectra emulators in the near future.

\section{Bispectra: effective formulas and halo model}
\label{sec-6}
We compute matter bispectra and bispectra between different tracers. For matter bispectra, we have implemented the standard perturbation theory Tree-level prediction as well as effective formulas from Ref. \cite{Scoccimarro:2000ee}  and Ref. \cite{Gil_Mar_n_2012} to predict the bispectrum for non-linear scales. We have also implemented the halo model bispectrum, predicting separately the 1, 2 and 3-halo terms. Currently implemented halo model bispectra include the matter bispectrum, the tSZ bispectrum and hybrid bispectra involving kSZ, tSZ, galaxies, and CMB weak lensing. 

\section{Halo mass function, halo bias}
\label{sec-6}
Predictions for halo model observables are ensemble averages of radial tracer profiles over halo masses and redshifts. The abundance of halos as a function of mass and redshift is characterized by the halo mass function. In \texttt{class\_sz} we have implemented several halo mass functions including the widely used Tinker formulas \cite{Tinker:2008ff,Tinker_2010}, as well as other fitting formulas \cite{Bocquet:2015pva}. Other mass functions based on emulators are currently being implemented\footnote{see, e.g., \href{https://github.com/SebastianBocquet/MiraTitanHMFemulator}{https://github.com/SebastianBocquet/MiraTitanHMFemulator}}. 
Similarly, we have fitting formulas for the first and second order biases following Ref. \cite{Tinker_2010} and Ref. \cite{Hoffmann:2015mma}, respectively. 

\section{Scale dependent bias from primordial non-Gaussianity}
It is well-known that local primordial non-Gaussianity generates a scale dependence in the halo bias and that the effect is  larger on large scales.  \texttt{class\_sz} can compute this scale-dependent effect following the standard calculation (see Ref. \cite{Dalal:2007cu}). If the user requests $f_{NL}\neq 0$, the effect is propagated to all calculations that depend on the halo bias.

\begin{figure}
\centering
\includegraphics[scale=0.35]{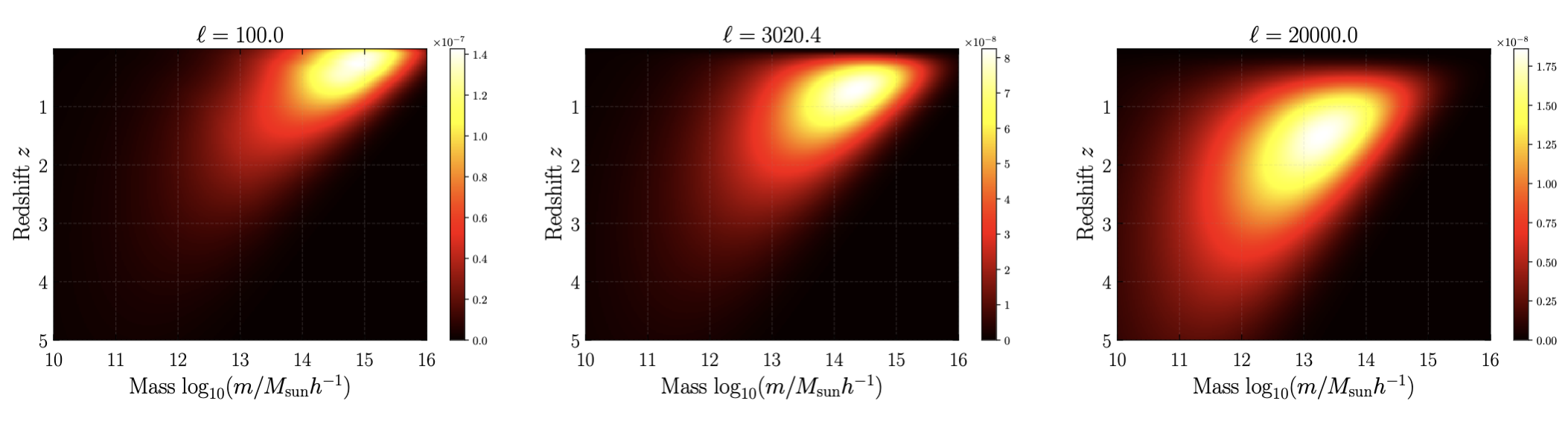}
\caption{Integrand of the angular power spectrum of the tSZ effect for $\ell=100$ (left), 3020.4 (center) and 20000 (right). The color scale indicates the integrand values. We see that at $\ell\approx3000$ the tSZ power spectrum is mainly sourced by halos of 10$^{14}$ M$_\mathrm{sun}/h$ and at $z\approx1$. See \href{https://github.com/CLASS-SZ/notebooks/blob/main/class_sz_szkernel.ipynb}{class\_sz\_szkernel.ipynb}.}
\vspace{-0.5cm}
\label{fig-1}       
\end{figure}

\section{Thermal SZ}

\texttt{class\_sz} uses the halo-model to compute auto- and cross-power spectra involving the thermal Sunyaev--Zeldovich (tSZ) effect. Several pressure profile models are implemented, including the Arnaud \emph{et al.}. \cite{Arnaud_2010} and the Battaglia \emph{et al.} \cite{2012ApJ...758...75B} generalized Navarro Frenk White (NFW) fitting formulas. One can request the radial pressure profile as a function of radius, or integrated quantities such as the mean Compton-$y$ parameter relevant to CMB spectral distortion. The thermal SZ power spectrum was the first observable to be implemented in \texttt{class\_sz} and is described in detail in Ref. \cite{Bolliet:2017lha}. We show the integrand of $C_\ell^{yy}$ in Figure \ref{fig-1}.
It is also possible to compute the tSZ power spectrum corresponding to a sub sample of halos determined by a selection function as was done in Ref. \cite{Rotti:2020rdl}, which can shed light on the mass-dependence of the hydrostatic mass bias and the possible role of relativistic SZ \cite{Remazeilles2019rSZ}. We note that the \texttt{class\_sz} Compton-$y$ calculations were used to benchmark the \texttt{ccl} implementation.

\section{Kinetic SZ}
 For the kinetic Sunyaev--Zeldovich (kSZ) effect, various gas density profiles that are currently available in \texttt{class\_sz} include a simple NFW scaled by the baryon fraction, as well as the more realistic Battaglia \emph{et al.} \cite{Battaglia:2016xbi} (gNFW) and Schneider \emph{et al.} \cite{Schneider:2015wta} (Baryonic Correction Model) formulas that were directly fitted to hydrodynamical simulations. From these profiles, one can then compute the contributions to the angular power spectra of the kSZ effect based on a formula valid at small scales (see Ref. \cite{Bolliet:2022pze} for details). 

\section{Galaxy power spectra, shear and intrinsic alignment}

Galaxy power spectra can be computed either within a linear bias model, e.g., $P_{gg}=b_g^2 P_L$, or from Halo Occupation Distributions. The HODs currently available are described in Ref. \cite{Kusiak:2022xkt} where they were used with \texttt{class\_sz} to characterize the galaxy-halo connection of the \textit{unWISE} galaxies. Galaxy lensing magnification is implemented too (see Ref. \cite{Kusiak:2022xkt}). 
We compute galaxy weak lensing power spectra, either in the linear bias approximation or within the halo model, and if requested, convert them to shear correlation functions in position space using \texttt{mcfit}\footnote{\href{https://github.com/eelregit/mcfit}{https://github.com/eelregit/mcfit}} Melin transforms \cite{2019ascl.soft06017L}. Our \texttt{class\_sz} shear calculations were benchmarked against those presented in Ref. \cite{DES:2021olg}.
Similarly, galaxy angular power spectra can be converted into clustering 2-point correlation functions so that \texttt{class\_sz} can be used to perform 3x2 analyses as in Ref. \cite{DES:2021wwk}. For intrinsic alignment we have a simple Non Linear Alignment model that follows Ref. \cite{DES:2021sgf}. More intrinsic alignment models including TATT and halo-models will become available in the near future. 

\section{Cosmic infrared background}

We have two halo models for the CIB: the Shang \emph{et al.}  \cite{Shang_2012} and the Maniyar \emph{et al.} \cite{Maniyar:2020tzw} models. The Shang \emph{et al.} model was used in, e.g., the \emph{Planck} 2013 CIB paper \cite{Planck:2013cib}, the \texttt{Websky} simulations \cite{Stein:2020its}, and in Ref. \cite{McCarthy:2020qjf}. The Maniyar \emph{et al.} model implementation was carefully benchmarked against the original code\footnote{\href{https://github.com/abhimaniyar/halomodel_cib_tsz_cibxtsz}{https://github.com/abhimaniyar/halomodel\_cib\_tsz\_cibxtsz}}. We can compute auto- and cross-frequency power spectra at any frequency, as well as prediction for the CIB monopole. Ref \cite{Sabyr:2022lhj} have used the CIB monopole implementation to predict the distortion caused by inverse Compton scattering of CIB photons against hot ICM electrons, see Ref.\cite{Sabyr:2022lhj}, although not accounting for intra-cluster scattering \cite{Acharya:2022xgn}.

\section{Cross-correlations}

All the tracers described above can be cross-correlated and their cross-power spectra can be computed with \texttt{class\_sz}. For instance, the cross-power spectrum between tSZ and CMB lensing measured from \textit{Planck} PR4 data was analysed with \texttt{class\_sz} in Ref. \cite{McCarthy:2023cwg}. Another recent work that relied on \texttt{class\_sz} for cross-correlation modelling is Ref. \cite{kusiak2023enhancing} where the authors proposed new component separation methods that incorporate information from cross-correlations between galaxies, CIB, and tSZ to better extract the CMB from multi-frequency maps. As a last example, a unique feature of \texttt{class\_sz} is the prediction for projected-field kSZ$^2$-LSS power spectra (we refer to Ref. \cite{Bolliet:2022pze} for details).

\section{SZ cluster counts}

With precise weak lensing mass calibration that will be enabled by VRO and \emph{Euclid} data, we expect promising constraints on the fundamental parameters of the universe from SZ selected cluster cosmology. Given a survey noise map and completeness function, \texttt{class\_sz} can predict the abundance of SZ selected clusters in SZ signal-to-noise and redshift bins. Scaling relations and completeness functions for \textit{Planck} (see Ref. \cite{Bolliet_2020} for a cluster cosmology analysis with \texttt{class\_sz}), ACT and SO-like surveys are already implemented. In addition to predicting the cluster abundance, if a catalogue data is assumed, \texttt{class\_sz} can also compute binned (see, e.g., Ref. \cite{plc_cc2016}) and unbinned (see, e.g., Ref. \cite{2019ApJ...878...55B}) likelihood values. The unbinned likelihood was developed in parallel to a new code called \texttt{cosmocnc} (to be released soon), specifically dedicated to cluster cosmology and which, for this particular purpose, is more general than \texttt{class\_sz}.

\section{Make it your own}
 For those interested in adding a new tracer profile that is not currently implemented, we have added a \texttt{custom\_profile} option which allows the user to pass a radial and a redshift kernel that is then automatically passed to \texttt{class\_sz} (see \href{https://github.com/CLASS-SZ/notebooks/blob/main/class_sz_tutorial.ipynb}{class\_sz\_tutorial.ipynb}). We note that this calculation is currently not parallelized and is therefore slower than native \texttt{class\_sz} calculations. This will be improved in the near future.

\KOMAoptions{fontsize=8pt}

\section*{Acknowledgements}

BB is grateful to D. Alonso, W. Coulton, E. Komatsu, J. Lesgourgues, A. Lewis and O. Philcox for very useful inputs. BB acknowledges  support from the European Research Council (ERC) under the European Union’s Horizon 2020 research and innovation programme (Grant agreement No. 851274).



%
\renewcommand*{\bibfont}{\footnotesize}
\bibliography{refs}

%
%
%
%




\end{document}